\documentclass{llncs}

\usepackage[T1]{fontenc}
\usepackage[utf8]{inputenc}
\usepackage{amsmath}
\usepackage{graphicx}
\usepackage{subfig}
\usepackage{color}

\newcommand{\stare}[1]{}
\newcommand{\nowe}[1]{#1}

\begin{document}
\title{MSARC: Multiple Sequence Alignment by Residue Clustering}
\author{Michał Modzelewski \and Norbert Dojer}
\institute{Insitute of Informatics, University of Warsaw, Poland\\
\email{dojer@mimuw.edu.pl}}

\maketitle              

\begin{abstract}

Progressive methods offer efficient and reasonably good solutions 
to the multiple sequence alignment problem.
However, resulting alignments are biased by guide-trees, 
especially for relatively distant sequences.

We propose MSARC, a new graph-clustering based algorithm
that aligns sequence sets without guide-trees.
Experiments on the BAliBASE dataset show that 
MSARC achieves alignment quality similar to best progressive methods
and substantially higher than the quality of other non-progressive algorithms.
\stare{Furthermore, MSARC outperforms all other methods on sequence sets 
with the similarity structure hardly represented by a phylogenetic tree.}
\nowe{Furthermore, MSARC outperforms all other methods on sequence sets 
whose evolutionary distances are hardly representable by a phylogenetic tree.}
These datasets are most exposed to the guide-tree bias of alignments.

MSARC is available at \url{http://bioputer.mimuw.edu.pl/msarc}

\end{abstract}

\section{Introduction}

Determining the alignment of a group of biological sequences
is among the most common problems in computational biology.
The dynamic programming method of pairwise sequence alignment 
can be readily extended to multiple sequences 
but requires the computation of an $n$-dimensional matrix to align $n$ sequences. 
Consequently, this method has an exponential time and space complexity. 

\emph{Progressive alignment} \cite{Thompson1994} offers 
a substantial complexity reduction 
at the cost of possible loss of the optimal solution.
Within this approach, subset alignments 
are sequentially pairwise aligned to build the final multiple alignment. 
The order of pairwise alignments is determined by
a guide-tree representing the phylogenetic relationships between sequences.

There are two drawbacks of the progressive alignment approach.
First, the accuracy of the guide-tree affects the quality of the final alignment.
This problem is particularly important in the field of phylogeny reconstruction,
because multiple alignment acts as a preprocessing step 
in most prominent methods of inferring a phylogenetic tree of sequences.
It has been shown that, within this approach, 
the inferred phylogeny is biased towards the initial guide-tree
\cite{Wong2008,Loeytynoja2008a}.

Second, only sequences belonging to currently aligned subsets 
contribute to their pairwise alignment.
Even if a guide-tree reflects correct phylogenetic relationships, 
these alignments may be inconsistent with remaining sequences
and the inconsistencies are propagated to further steps.
To address this problem, in recent programs 
\cite{Notredame2000,Edgar2004a,Katoh2005a,Do2005,Roshan2006}
progressive alignment is usually preceded by \emph{consistency transformation} 
(incorporating information from all pairwise alignments into the objective function)
and/or followed by \emph{iterative refinement} of 
the multiple alignment of all sequences.
 
In the present paper we propose MSARC, 
a new multiple sequence alignment algorithm that avoids guide-trees altogether.
MSARC constructs a graph with all residues from all sequences as nodes
and edges weighted with alignment affinities of its adjacent nodes.
Columns of best multiple alignments tend to form clusters in this graph,
so in the next step residues are clustered (see Figure~\ref{fig:overview}a).
Finally, MSARC refines the multiple alignment corresponding to the clustering.

Experiments on the BAliBASE dataset \cite{Thompson2005} show that our approach 
is competitive with the best progressive methods and 
significantly outperforms current non-progressive algorithms
\cite{Subramanian2005,Subramanian2008}.
Moreover, MSARC is the best aligner for sequence sets
with very low levels of conservation.
This feature makes MSARC a promising preprocessing tool 
for phylogeny reconstruction pipelines.

\section{Methods}

\begin{figure}[!t]
\begin{minipage}[t]{0.35\columnwidth}%

\begin{center}
(a)\includegraphics[clip,width=0.9\columnwidth]{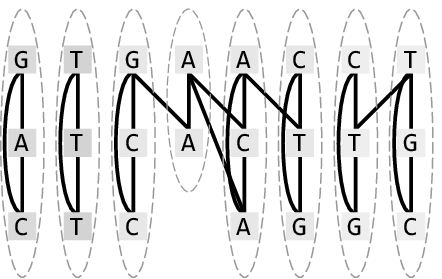}

\vspace{1ex}

\begin{tabular}{cccc}
(b)
&
{\includegraphics[clip,width=0.2\columnwidth]{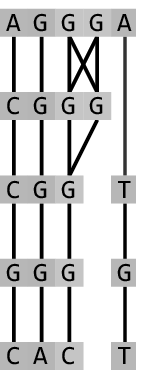}}
&
{\includegraphics[clip,width=0.2\columnwidth]{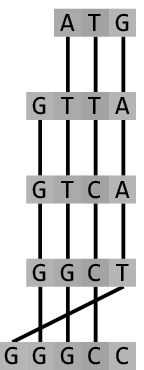}}
&
{\includegraphics[bb=0bp 0bp 40bp 104bp,clip,width=0.2\columnwidth]{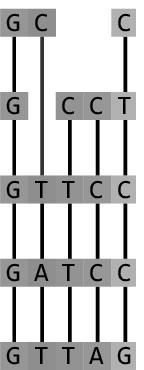}}
\\
&
Ambiguity
&
Conflict
&
Valid
\end{tabular}
\end{center}
\end{minipage}\hfill{}%
\begin{minipage}[t]{0.6\columnwidth}%
\vspace{-17ex}
(c)
\includegraphics[width=0.95\columnwidth]{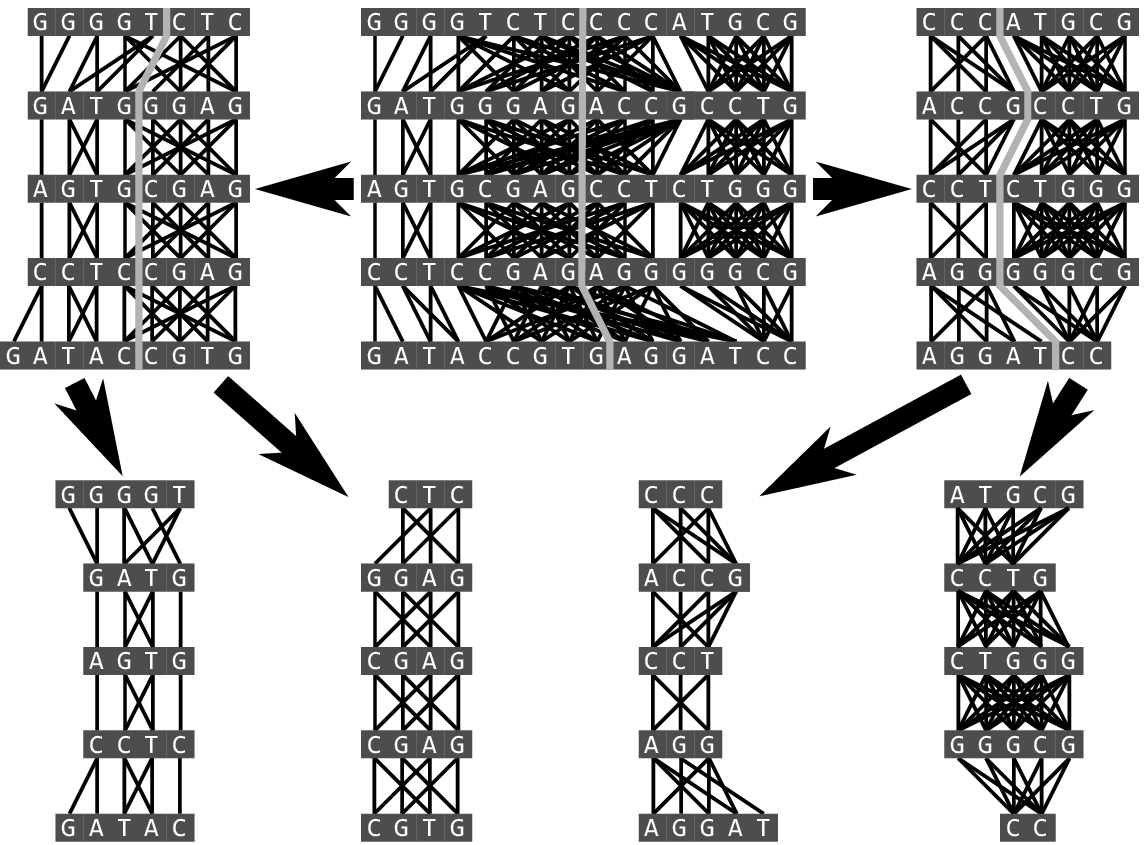}
\end{minipage}
\caption{Overview of our residue clustering approach.
(a) Alignment graph and its desired clustering.
Clusters form columns of a corresponding multiple sequence alignment. 
(b) Clusterings inconsistent (left and middle) and consistent (right) 
with the alignment structure. 
(c) An example of hierarchical divisive clustering of residues. 
The graph is recursively partitioned by finding a balanced minimal cut 
while maintaining the ordering of residues until
all parts have at most one residue from each sequence.
Final alignment is constructed by concatenating these parts (alignment columns)
from left to right.}\label{fig:overview}
\end{figure}

MSARC aligns sequence sets in several steps.
In a preprocessing step, following Probalign \cite{Roshan2006},
\emph{stochastic alignments} are calculated for all pairs of sequences
and consistency transformation is applied to resulting
posterior probabilities of residue correspondences.
Transformed probabilities, called residue alignment affinities,
represent weights of an \emph{alignment graph}\footnote{Our notion of
alignment graph slightly differs from the one of Kececioglu \cite{Kececioglu1993}:
removing edges between clusters transforms the former into the latter.}.
MSARC clusters this graph with a top-down hierarchical method (Figure~\ref{fig:overview}c).
Division steps are based on the Fiduccia-Mattheyses 
graph partitioning algorithm \cite{Fiduccia1982},
adapted to satisfy constraints imposed by the sequence order of residues.
Finally, multiple alignment corresponding to resulting clustering
is refined with the iterative improvement strategy
proposed in Probcons \cite{Do2005},
adapted to remove clustering artefacts.

\subsection{Pairwise stochastic alignment}

The concept of stochastic (or probability) alignment was proposed in \cite{Miyazawa1995}.
Given a pair of sequences, this framework defines 
statistical weights of their possible alignments.
Based on these weights, for each pair of residues from both sequences,
the posterior probability of being aligned may be computed. 
A consensus of highly weighted suboptimal alignments was shown 
to contain pairs with significant probabilities 
that agree with structural alignments 
despite the optimal alignment deviating significantly.
M\"uckstein et al. \cite{Mueckstein2002} suggest the use of the method as a starting point
for improved multiple sequence alignment procedures. 

The statistical weight $\mathcal{W}\left(\mathcal{A}\right)$ 
of an alignment $\mathcal{A}$
is the product of the individual weights of (mis-)matches and gaps \cite{Yu2001}.
It may be obtained from the standard similarity scoring function $S(\mathcal{A})$
with the following formula:
\begin{equation}
\mathcal{W}\left(\mathcal{A}\right)=e^{\beta{S\left(\mathcal{A}\right)}}
\end{equation}
\nowe{where $\beta$ corresponds to the inverse of Boltzmann's constant
and should be adjusted to the match/mismatch scoring function $s(x,y)$
(in fact, $\beta$ simply rescales the scoring function).}

The probability distribution over all alignments $\mathcal{A}^{*}$
is achieved by normalizing this value. The normalization factor $Z$
is called the \emph{partition function} of the alignment problem \cite{Miyazawa1995},
and is defined as
\begin{equation}
Z=\sum_{\mathcal{A}\in\mathcal{A}^{*}}\mathcal{W}\left(\mathcal{A}\right)=\mathcal{\sum_{\mathcal{A}\in\mathcal{A}^{*}}}e^{\beta{S\left(\mathcal{A}\right)}}
\end{equation}

The probability $P\left(\mathcal{A}\right)$ of an alignment can be calculated by
\begin{equation}
P\left(\mathcal{A}\right)=\frac{\mathcal{W}\left(\mathcal{A}\right)}{Z}
=\frac{e^{\beta{S\left(\mathcal{A}\right)}}}{Z}
\end{equation}

Let $\mathbf{P}\left(a_{i}\sim b_{j}\right)$ denote
the posterior probability that residues $a_{i}$ and $b_{j}$ are aligned.
We can calculate it as the sum of probabilities of all alignments 
with $a_{i}$ and $b_{j}$ in a common column (denoted by ${A}^{*}_{a_{i}\sim b_{j}}$):
\begin{multline}
\mathbf{P}\left(a_{i}\sim b_{j}\right) 
 = \sum_{\mathcal{A}\in\mathcal{A}^{*}_{a_{i}\sim b_{j}}\hspace{-4ex}}
 \mathbf{P}(\mathcal{A})
 = \frac{{\displaystyle 
 \sum_{\mathcal{A}\in\mathcal{A}^{*}_{a_{i}\sim b_{j}}\hspace{-4ex}}}
 e^{\beta S\left(\mathcal{A}\right)}}{Z} =\\
 = \frac{\displaystyle \bigg(\sum_{\mathcal{A}_{i-1,j-1}\hspace{-5ex}}
   e^{\beta S(\mathcal{A}_{i-1,j-1})}\bigg)
  e^{\beta s(a_{i},b_{j})}
  \bigg(\sum_{\widehat{\mathcal{A}}_{i+1,j+1}\hspace{-5ex}}
   e^{\beta S(\widehat{\mathcal{A}}_{i+1,j+1})}\bigg)}
  {Z} =\\
 = \frac{Z_{i-1,j-1}\, e^{\beta s\left(a_{i},b_{j}\right)}\,\widehat{Z}_{i+1,j+1}}
  {Z}
\end{multline}
Here we use the notation ${\textstyle \mathcal{A}}_{i,j}$ for an
alignment of the sequence prefixes $a_{1}\cdots a_{i}$ and $b_{1}\cdots b_{j}$,
and $\widehat{{\textstyle \mathcal{A}}}_{i,j}$ for an alignment of
the sequence suffixes $a_{i}\cdots a_{m}$ and $b_{j}\cdots b_{n}$.
Analogously, $Z_{i,j}$ is the partition function over the prefix alignments
and $\widehat{Z}_{i,j}$ is the (reverse) partition function over
the suffix alignments. 

\nowe{An efficient algorithm for calculating the partition function can
be derived from the Gotoh maximum score algorithm \cite{Gotoh1982} 
by replacing the maximum operations with additions.
From a few possible approaches \cite{Miyazawa1995,Yu2001,Mueckstein2002}
we chose a variant proposed by Miyazawa \cite{Miyazawa1995} 
and applied in Probalign \cite{Roshan2006},
where insertions and deletions must be separated 
by at least one match/mismatch position:}
\begin{eqnarray}
Z_{i,j}^{M} & = & \left(Z_{i-1,j-1}^{M}+Z_{i-1,j-1}^{E}+Z_{i-1,j-1}^{F}\right)e^{\beta s\left(a_{i},b_{j}\right)}\\
Z_{i,j}^{E} & = & Z_{i,j-1}^{M}e^{\beta g_{o}}+Z_{i,j-1}^{E}e^{\beta g_{ext}}\label{eq:partition-e}\\
Z_{i,j}^{F} & = & Z_{i-1,j}^{M}e^{\beta g_{o}}+Z_{i-1,j}^{F}e^{\beta g_{ext}}\\
Z_{i,j} & = & Z_{i,j}^{M}+Z_{i,j}^{E}+Z_{i,j}^{F}
\end{eqnarray}
The reverse partition function can be calculated using the same recursion
in reverse, starting from the ends of the aligned sequences.

\subsection{Alignment graphs}

Probabilities
$\mathbf{P}\left(a_{i}\sim b_{j}\right)$ 
may be viewed as a representation of a bipartite graph
with nodes corresponding to residues $a_i$ and $b_j$
and edges weighted with residue alignment affinity.

Given a set $S$ of $k$ sequences to be aligned,
we would like to analogously represent their residue alignment affinity
by a $k$-partite weighted graph.
It may be obtained by joining pairwise alignment graphs
for all pairs of $S$-sequences.
However, separate computation of edge weights for each pair of sequences
does not exploit information included in the remaining alignments.
In order to incorporate correspondence with residues from other sequences, 
we perform a \emph{consistency transformation} \cite{Notredame2000,Do2005}.
It re-estimates the residue alignment affinity 
according to the following formula:
\nowe{\begin{equation}
\mathbf{P}^{\prime}\left(x_{i}\sim y_{j} \right) 
\leftarrow \frac{{\displaystyle \sum_{z\in S}}{\displaystyle \sum_{l=0}^{|z|}}
\mathbf{P}\left(x_{i}\sim z_{l}\right)
\mathbf{P}\left(z_{l}\sim y_{j}\right)}{\left|S\right|}
\end{equation}}
If $P_{xy}$ is a matrix of current residue alignment affinities
for sequences $x$ and $y$, the matrix form equivalent
transformation is
\begin{equation}
P_{xy}^{\prime}\leftarrow\frac{{\displaystyle \sum_{z\in S}}P_{xz}P_{zy}}{\left|S\right|}\label{eq:consistency-matrix}
\end{equation}

\nowe{The consistency transformation may be iterated any number of times,
but excessive iterations blur the structure of residue affinity.
Following Probalign \cite{Roshan2006} and ProbCons \cite{Do2005} 
MSARC performs it twice by default.}

\subsection{Residue clustering}

Columns of any multiple alignment form a partition of the set of sequence residues.
The main idea of MSARC is to reconstruct the alignment 
by clustering an alignment graph into columns.
The clustering method must satisfy constraints imposed by alignment structure.
First, each cluster may contain at most one residue from a single sequence.
Second, the set of all clusters must be orderable
consistently with sequence orders of their residues.
Violation of the first constraint will be called \emph{ambiguity}, 
while violation of the second one -- \emph{conflict} 
(see Figure~\ref{fig:overview}b).

Towards this objective, MSARC applies top-down hierarchical clustering
(see Figure~\ref{fig:overview}c).
Within this approach, the alignment graph is recursively split into two parts
until no ambiguous cluster is left.
Each partition step results from a single cut through all sequences,
so clusterings are conflict-free at each step of the procedure.
Consequently, the final clustering represents a proper multiple alignment.

%

Optimal clustering is expected 
to maximize residue alignment affinity within clusters
and minimize it between them.
Therefore, the partition selection in recursive steps of the clustering procedure
should minimize the sum of weights of edges cut by the partition.
This is in fact the objective of the well-known problem of \emph{graph partitioning},
i.e. dividing graph nodes into roughly equal parts such that 
the sum of weights of edges connecting nodes in different parts is minimized.

The Fiduccia-Mattheyses algorithm \cite{Fiduccia1982} is an efficient
heuristic for the graph partitioning problem.
After selecting an initial, possibly random partition, it calculates
for each node the change in cost caused by moving it between parts, 
called \textit{gain}. 
Subsequently, single nodes are greedily moved between partitions based on the maximum gain
and gains of remaining nodes are updated. 
The process is repeated in \textit{passes},
where each node can be moved only once per pass. 
The best partition found in a pass is chosen 
as the initial partition for the next pass. 
The algorithm terminates when a pass fails to improve the partition. 
\nowe{Grouping single moves into passes helps the algorithm to escape local optima,
since intermediate partitions in a pass may have negative gains.}
An additional balance condition is enforced, 
disallowing movement from a partition that contains
less than a minimum desired number of nodes.

Fiduccia-Mattheyses algorithm needs to be modified
in order to deal with alignment graphs. 
Mainly, residues are not moved independently; 
since the graph topology has to be maintained, 
moving a residue involves moving all the residues 
positioned between it and a current cut point on its sequence.
This modification implies further changes 
in the design of data structures for gain processing.
\nowe{Next, the sizes of parts in considered partitions cannot differ
by more than the maximum cluster size in a final clustering, 
i.e., the number of aligned sequences.}
This choice implies minimal search space containing partitions 
consistent with all possible multiple alignment.
In the initial partition sequences are cut in their midpoints.

%

The Fiduccia-Mattheyses heuristic may be optionally extended
with a \emph{multilevel} scheme \cite{Hendrickson1995}.
In this approach increasingly coarse approximations of the graph 
are created by an iterative process called \emph{coarsening}. 
At each iteration step selected pairs of nodes are merged into single nodes. 
Adjacent edges are merged accordingly and weighted with sums of original weights.
The final coarsest graph is partitioned using Fiduccia-Mattheyses algorithm.
Then the partition is projected back to the original graph 
through the series of \emph{uncoarsening} operations, 
each of which is followed by a Fiduccia-Mattheyses based refinement.
Because the last refinement is applied
to the original graph, the multilevel scheme in fact 
reduces the problem of selecting an initial partition
to the problem of selecting pairs of nodes to be merged.
In alignment graphs only neighboring nodes can be merged, so
MSARC just merges consecutive pairs of neighboring nodes.

\subsection{Refinement}

\begin{figure}[tbh]
\begin{center}
\includegraphics[height=6ex]{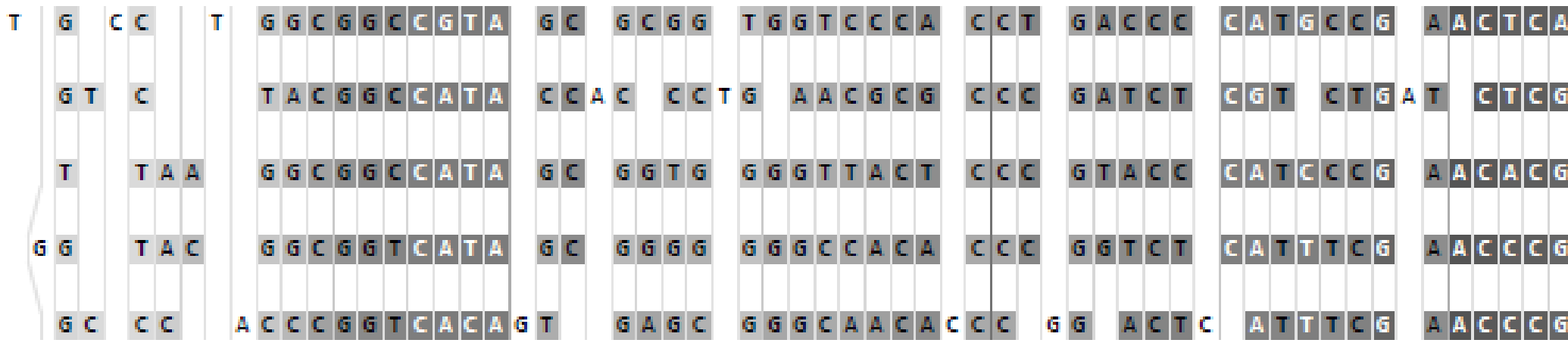}\\
(a) Fiduccia-Mattheyses partitioning\\

\includegraphics[height=6ex]{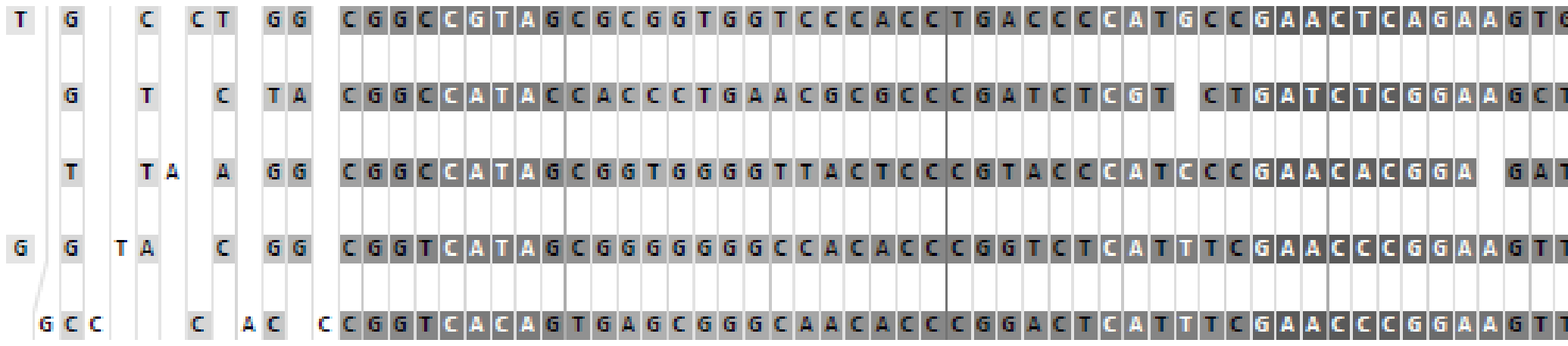}\\
(b) Multilevel partitioning\\

\includegraphics[height=6ex]{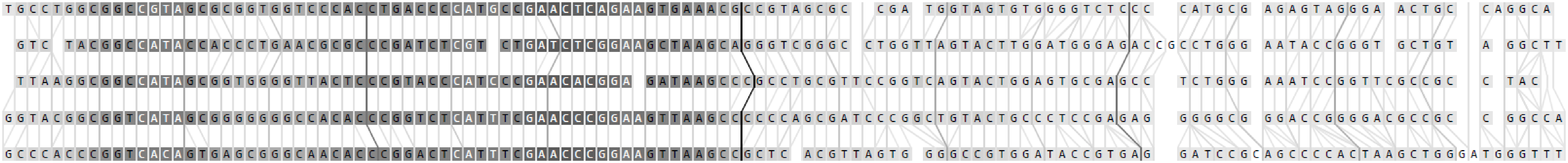}\\
(c) Refined Fiduccia-Mattheyses partitioning\\

\includegraphics[height=6ex]{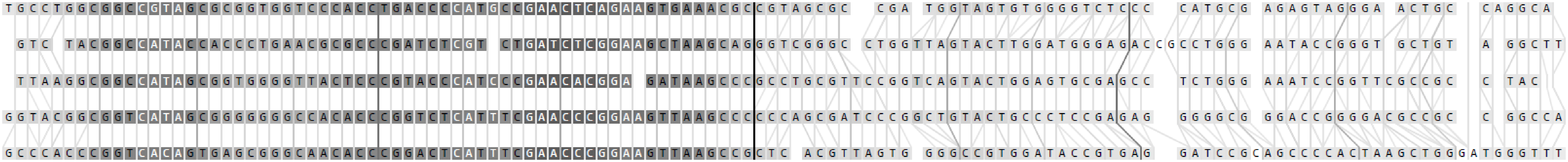}\\
(d) Refined multilevel partitioning\\
\end{center}

\caption{\nowe{Example visualization of the alignment produced by the graph partitioning
methods alone (ab) and graph partitioning followed by refinement (cd). 
Residue colors reflect how well the column is aligned 
based on residue match probabilities (darker is better).
Partition cuts are colored to show the order of partitioning 
with darker cuts being performed earlier. 
}}
\label{fig:part_example}
\end{figure}

\nowe{An example of alignment columns produced by residue clustering can be
seen in Figure \ref{fig:part_example}(ab).
Unfortunately, right parts of alignments contain many superfluous spaces
that could easily be removed manually.}

\nowe{Therefore we decided to add a refinement step, 
following the method used in ProbCons \cite{Do2005}.}
Sequences are split into two groups and the groups are pairwise re-aligned.
Re-alignment is performed using the Needleman-Wunsch algorithm
with the score for each pair of positions defined as the sum of posterior
probabilities for all non-gap pairs and zero gap-penalty. 
Since gap-penalties are not used, 
every such refinement iteration creates a new alignment of equal or greater expected accuracy. 
\nowe{First each sequence is re-aligned with the remaining sequences,
since such division is very efficient in removing superfluous spaces.} 
Next, several randomly selected sequence subsets are re-aligned against the rest.

\nowe{Figures \ref{fig:part_example}(cd) show the results 
of refining the alignments from Figures \ref{fig:part_example}(ab). 
Refinement removed superfluous spaces from the clustering process 
and optimized the alignment. 
Note that the final post-refinement alignments turned out to be the same for both
Fiduccia-Mattheyses and multilevel method of graph partitioning.}

%

\section{Results}

\subsection{Benchmark data and methodology}

MSARC was tested against the BAliBASE
3.0 benchmark database \cite{Thompson1994}. 
It contains manually
refined reference alignments based on 3D structural superpositions.
Each alignment contains core-regions that correspond to the most reliably
alignable sections of the alignment. Alignments are divided into five
sets designed to evaluate performance on varying types of problems: 
\begin{itemize}
\item[{\textsc{rv1x}}] Equidistant sequences with two different levels
of conservation 

\begin{itemize}
\item[{\textsc{rv11}}] very divergent sequences (<20\% identity)
\item[{\textsc{rv12}}] medium to divergent sequences (20-40\% identity)
\end{itemize}
\item[{\textsc{rv20}}] Families aligned with a highly divergent ``orphan''
sequence
\item[{\textsc{rv30}}] Subgroups with <25\% residue identity between groups
\item[{\textsc{rv40}}] Sequences with N/C-terminal extensions
\item[{\textsc{rv50}}] Internal insertions
\end{itemize}

%

BAliBASE 3.0 also provides a program 
comparing given alignments with a reference one.
Alignments are scored according to two metrics. A sum-of-pairs
score (SP) showing the ratio of residue pairs that are correctly
aligned, and a total column (TC) score showing the ratio
of correctly aligned columns. Both scores can be applied to
full sequences or just the core-regions. 

Two variants of MSARC: with multilevel Fiduccia-Mattheyses algorithm (MSARC-ML)
and with basic Fiduccia-Mattheyses algorithm (MSARC-FM)
were tested on the full length sequences
and scored based on the correct alignment of core-regions.
\nowe{The results were compared to 
CLUSTAL $\mathrm{\Omega}$ \cite{Thompson1994,Sievers2011} ver. 1.1.0,
DIALIGN-T \cite{Subramanian2005} ver. 0.2.2, 
DIALIGN-TX \cite{Subramanian2008} ver. 1.0.2,
MAFFT \cite{Katoh2005a} %
ver. 6.903, MUSCLE \cite{Edgar2004a} ver. 3.8.31, 
MSAProbs \cite{Liu2009} ver. 0.9.7, Probalign \cite{Roshan2006} ver. 1.4, 
ProbCons \cite{Do2005} ver. 1.12 and T-Coffee \cite{Notredame2000} ver. 9.02.}

\nowe{All the programs were executed with their default parameters.
In the case of MSARC, default parameters of \emph{stochastic alignment},
\emph{consistency transformation} and \emph{iterative refinement} steps
follow the defaults of corresponding steps of Probalign and ProbCons.
Namely, MSARC was run with 
Gonnet 160 similarity matrix \cite{Gonnet1992},
gap penalties of $-22$, $-1$ and $0$ for gap open, extension and
terminal gaps respectively, $\beta=0.2$, a
cut-off value for posterior probabilities of $0.01$
(values smaller than the cutoff are set to $0$ and
operations designed for sparse matrices are
used in order to speed up computations), 
two iterations of the consistency transformation and 
$100$ iterations of iterative refinement.}

\subsection{Aligner comparison}

\begin{table}[!t]
\caption{Performance on BAliBASE 3.0\label{tab:results}}
\begin{center}
{\begin{tabular}{llcccccccclr}
\hline 
 & & \multicolumn{8}{c}{SP/TC scores}  & & Computation  \\
{Aligner} & & {all} & \textsc{rv11} & \textsc{rv12} & \textsc{rv20} & \textsc{rv30} & \textsc{rv40} & \textsc{rv50} & \textsc{bb40037} & & {Time}\\
\hline \\[1ex]
MSARC-ML & & $\displaystyle\frac{87.6}{57.3}$ & $\displaystyle\frac{\mathbf{70.1}}{\mathbf{46.1}}$ & $\displaystyle\frac{94.5}{85.6}$ & $\displaystyle\frac{92.5}{40.7}$ & $\displaystyle\frac{83.4}{45.7}$ & $\displaystyle\frac{\mathbf{93.1}}{\mathbf{63.3}}$ & $\displaystyle\frac{88.7}{51.6}$ & $\displaystyle\frac{\mathbf{97.1}}{\mathbf{70.0}}$ & & ${33:49:37}$\\ \\[1ex] 
MSARC-FM & & $\displaystyle\frac{87.5}{57.1}$ & $\displaystyle\frac{70.0}{46.0}$ & $\displaystyle\frac{94.5}{85.6}$ & $\displaystyle\frac{92.5}{40.9}$ & $\displaystyle\frac{82.8}{45.0}$ & $\displaystyle\frac{93.0}{62.9}$ & $\displaystyle\frac{88.6}{51.7}$ & $\displaystyle\frac{\mathbf{97.1}}{\mathbf{70.0}}$ & & ${22:14:19}$\\ \\[1ex] 
{CLUSTAL $\mathrm{\Omega}$} & & $\displaystyle\frac{84.0}{55.4}$ & $\displaystyle\frac{59.0}{35.8}$ & $\displaystyle\frac{90.6}{78.9}$ & $\displaystyle\frac{90.2}{45.0}$ & $\displaystyle\frac{86.2}{57.5}$ & $\displaystyle\frac{90.2}{57.9}$ & $\displaystyle\frac{86.2}{53.3}$ & $\displaystyle\frac{61.2}{0.0}$ & & $\mathbf{12:15}$\\ \\[1ex] 
DIALIGN-T & & $\displaystyle\frac{77.3}{42.8}$ & $\displaystyle\frac{49.3}{25.3}$ & $\displaystyle\frac{88.8}{72.5}$ & $\displaystyle\frac{86.3}{29.2}$ & $\displaystyle\frac{74.7}{34.9}$ & $\displaystyle\frac{82.0}{45.2}$ & $\displaystyle\frac{80.1}{44.2}$ & $\displaystyle\frac{52.6}{0.0}$ & & ${1:13:21}$\\ \\[1ex] 
DIALIGN-TX & & $\displaystyle\frac{78.8}{44.3}$ & $\displaystyle\frac{51.5}{26.5}$ & $\displaystyle\frac{89.2}{75.2}$ & $\displaystyle\frac{87.9}{30.5}$ & $\displaystyle\frac{76.2}{38.5}$ & $\displaystyle\frac{83.6}{44.8}$ & $\displaystyle\frac{82.3}{46.6}$ & $\displaystyle\frac{52.8}{0.0}$ & & ${1:36:05}$\\ \\[1ex] 
MAFFT & & $\displaystyle\frac{86.7}{58.4}$ & $\displaystyle\frac{65.3}{42.8}$ & $\displaystyle\frac{93.6}{83.8}$ & $\displaystyle\frac{92.5}{44.6}$ & $\displaystyle\frac{85.9}{58.1}$ & $\displaystyle\frac{91.5}{59.0}$ & $\displaystyle\frac{90.1}{59.4}$ & $\displaystyle\frac{56.4}{0.0}$ & & ${54:04}$\\ \\[1ex] 
MUSCLE & & $\displaystyle\frac{81.9}{47.5}$ & $\displaystyle\frac{57.2}{31.8}$ & $\displaystyle\frac{91.5}{80.4}$ & $\displaystyle\frac{88.9}{35.0}$ & $\displaystyle\frac{81.4}{40.9}$ & $\displaystyle\frac{86.5}{45.0}$ & $\displaystyle\frac{83.5}{45.9}$ & $\displaystyle\frac{48.4}{0.0}$ & & ${23:32}$\\ \\[1ex] 
MSAProbs & & $\displaystyle\frac{\mathbf{87.8}}{\mathbf{60.7}}$ & $\displaystyle\frac{68.2}{44.1}$ & $\displaystyle\frac{94.6}{\mathbf{86.5}}$ & $\displaystyle\frac{\mathbf{92.8}}{\mathbf{46.4}}$ & $\displaystyle\frac{\mathbf{86.5}}{\mathbf{60.7}}$ & $\displaystyle\frac{92.5}{62.2}$ & $\displaystyle\frac{\mathbf{90.8}}{\mathbf{60.8}}$ & $\displaystyle\frac{59.5}{0.0}$ & & ${6:43:51}$\\ \\[1ex] 
{Probalign} & & $\displaystyle\frac{87.6}{58.9}$ & $\displaystyle\frac{69.5}{45.3}$ & $\displaystyle\frac{\mathbf{94.6}}{86.2}$ & $\displaystyle\frac{92.6}{43.9}$ & $\displaystyle\frac{85.3}{56.6}$ & $\displaystyle\frac{92.2}{60.3}$ & $\displaystyle\frac{88.7}{54.9}$ & $\displaystyle\frac{54.2}{0.0}$ & & ${4:31:41}$\\ \\[1ex] 
{ProbCons} & & $\displaystyle\frac{86.4}{55.8}$ & $\displaystyle\frac{67.0}{41.7}$ & $\displaystyle\frac{94.1}{85.5}$ & $\displaystyle\frac{91.7}{40.6}$ & $\displaystyle\frac{84.5}{54.4}$ & $\displaystyle\frac{90.3}{53.2}$ & $\displaystyle\frac{89.4}{57.3}$ & $\displaystyle\frac {59.3}{0.0}$ & & ${6:56:32}$\\ \\[1ex] 
{T-Coffee} & & $\displaystyle\frac{85.7}{55.1}$ & $\displaystyle\frac{65.5}{40.9}$ & $\displaystyle\frac{93.9}{84.8}$ & $\displaystyle\frac{91.4}{40.1}$ & $\displaystyle\frac{83.7}{49.0}$ & $\displaystyle\frac{89.2}{54.5}$ & $\displaystyle\frac{89.4}{58.5}$ & $\displaystyle\frac{50.9}{0.0}$ & & ${13:53:02}$\\ \\[1ex] 
\hline
\end{tabular}}
\end{center}
\nowe{{Columns 2-9 show the mean SP and TC scores for each alignment
algorithm on the whole BAliBASE dataset,
each of its series and case \textsc{bb40037}. 
The last column presents total CPU
computation time (hh:mm:ss). 
All scores are multiplied by 100. 
Best results in each column are shown in bold.}}
\end{table}

Table~\ref{tab:results} shows the SP and TC
scores obtained by the alignment algorithms on 
the BAliBASE 3.0 benchmark. 
MSARC-ML has slightly better accuracy than MSARC-FM.
Both variants of MSARC substantially outperform DIALIGN-T 
(the only non-progressive method in the test) and DIALIGN-TX 
(a progressive extension of DIALIGN-T).
Moreover, MSARC achieves accuracy similar to the leading alignment methods:
MSAProbs, Probalign and ProbCons.

\begin{table}[!t]
\caption{Significance of differences in BAliBASE 3.0
 SP/TC scores\label{tab:p-values}}
\begin{center}
{\begin{tabular}{lccccccc}
\hline
{SP scores} & \textsc{rv11} & \textsc{rv12} & \textsc{rv20} & \textsc{rv30} & \textsc{rv40} & \textsc{rv50} & {Total}\\
\hline
{Clustal $\mathrm{\Omega}$} & {+3.8e-7} & {+1.1e-5} & {+0.0031} & {-0.047} & {+4.2e-6} & {+0.012} & {+8.7e-15}\\
DIALIGN-T & {+8.6e-8} & {+7.7e-9} & {+1.3e-7} & {+2.7e-6} & {+2.1e-9} & {+0.00098} & {+5.3e-36}\\
DIALIGN-TX & {+1.0e-7} & {+6.2e-8} & {+2.3e-7} & {+8.7e-6} & {+2.8e-9} & {+0.0017} & {+3.1e-34}\\
MAFFT & {+0.0031} & {+0.00085} & {-(0.64)} & {-0.0009} & {+0.0005} & {-(0.072)} & {+0.028}\\
MUSCLE & {+4.5e-6} & {+1.3e-6} & {+0.0002} & {+(0.24)} & {+2.5e-8} & {+0.006} & {+6.8e-22}\\
MSAProbs & {+0.015} & {-(0.56)} & {-0.016} & {-1.9e-5} & {+(0.39)} & {-0.0041} & {-0.0025}\\
{Probalign} & {+(0.16)} & {-(0.77)} & {-0.048} & {-0.0099} & {+(0.66)} & {-(0.85)} & {-(0.067)}\\
{ProbCons} & {+0.0070} & {+0.037} & {+0.032} & {-(0.11)} & {+0.0014} & {-(0.17)} & {+0.0018}\\
{T-Coffee} & {+0.001} & {+0.005} & {+0.021} & {-(0.40)} & {+0.0001} & {-(0.077)} & {+7.1e-6}\\
\hline
{TC scores} & \textsc{rv11} & \textsc{rv12} & \textsc{rv20} & \textsc{rv30} & \textsc{rv40} & \textsc{rv50} & {Total}\\
\hline
{Clustal $\Omega$} & {+2.8e-5} & {+0.0004} & {-0.025} & {-0.0018} & {+(0.11)} & {-(0.84)} & {+(0.096)}\\
DIALIGN-T & {+1.5e-6} & {+2.2e-8} & {+9.6e-5} & {+0.0024} & {+4.9e-8} & {+0.027} & {+3.6e-26}\\
DIALIGN-TX & {+1.3e-6} & {+4.0e-7} & {+0.00040} & {+0.038} & {+1.3e-7} & {+(0.066)} & {+9.5e-23}\\
MAFFT & {+(0.11)} & {+0.005} & {-(0.052)} & {-0.0007} & {+(0.07)} & {-(0.062)} & {-(0.55)}\\
MUSCLE & {+9.9e-5} & {+0.0002} & {+(0.06)} & {+(0.76)} & {+2.2e-6} & {+0.009} & {+5.8e-13}\\
MSAProbs & {+(0.13)} & {-(0.22)} & {-0.0016} & {-8.5e-5} & {+(0.076)} & {-0.0014} & {-5.4e-7}\\
{Probalign} & {+(0.54)} & {-(0.11)} & {-0.00062} & {-0.0006} & {+(0.087)} & {-(0.36)} & {-1.9e-6}\\
{ProbCons} & {+0.043} & {-(0.69)} & {-(0.31)} & {-0.011} & {+0.017} & {-(0.062)} & {+(0.84)}\\
{T-Coffee} & {+0.003} & {+(0.10)} & {+(0.75)} & {-(0.11)} & {+(0.12)} & {-0.0072} & {+(0.61)}\\
\hline
\end{tabular}}
\end{center}
{{
Entries show $p$-values indicating the significance of the mean difference of
SP/TC scores between MSARC-ML and other aligners 
as measured using the Wilcoxon matched-pair signed-rank test. 
A $+$ means that MSARC had a higher mean score 
while a $-$ means MSARC had a lower mean score. 
Nonsignificant $p$-values (>0.05) are shown in parentheses.}}
\end{table}

The differences are not significant in most cases 
(see Table~\ref{tab:p-values}) and
correspond with the structure of benchmark series --
MSARC shows the best
results for test series \textsc{rv11} and \textsc{rv40}, and the worst
performance on \textsc{rv20} and \textsc{rv30}. 
Distances in \textsc{rv20} and \textsc{rv30} families 
are particularly well represented by phylogenetic trees 
(low similarity between highly conserved subgroups).
On the other hand, series \textsc{rv11}
contains highly divergent sequences for which guide-tree is poorly informative,
even if it represents the correct phylogeny, 
and \textsc{rv40} contains sequences
with N/C-terminal extensions which may affect the accuracy
of the estimated phylogeny. 

\begin{figure}[!t]
\begin{center}
\begin{tabular}{lc}
\begin{tabular}{l}
BAliBASE
\\\\(a)
\end{tabular}
 & 
{\includegraphics[bb=-2500bp 0bp 9084bp 368bp,width=.75\columnwidth,height=0.2\columnwidth]{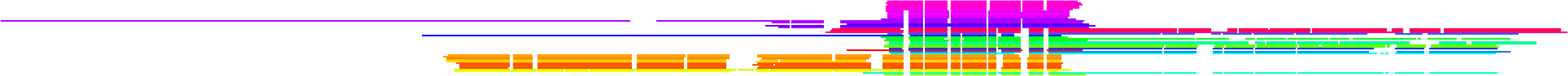}
}\\
\begin{tabular}{l}
MSARC
\\\\(b)
\end{tabular}
& 
{\includegraphics[bb=-2200bp 0bp 9384bp 368bp,width=.75\columnwidth,height=0.2\columnwidth]{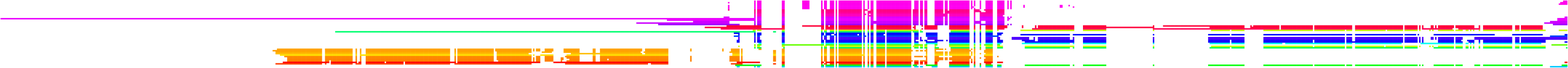}
}\\
\begin{tabular}{l}
Probalign
\\\\(c)
\end{tabular}
& 
{\includegraphics[bb=-2500bp 0bp 9084bp 368bp,width=.75\columnwidth,height=0.2\columnwidth]{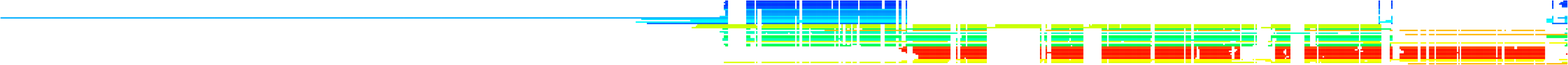}
}\\
\begin{tabular}{l}
MSAProbs
\\\\(d)
\end{tabular}
& 
{\includegraphics[width= .75\columnwidth,height=0.2\columnwidth]{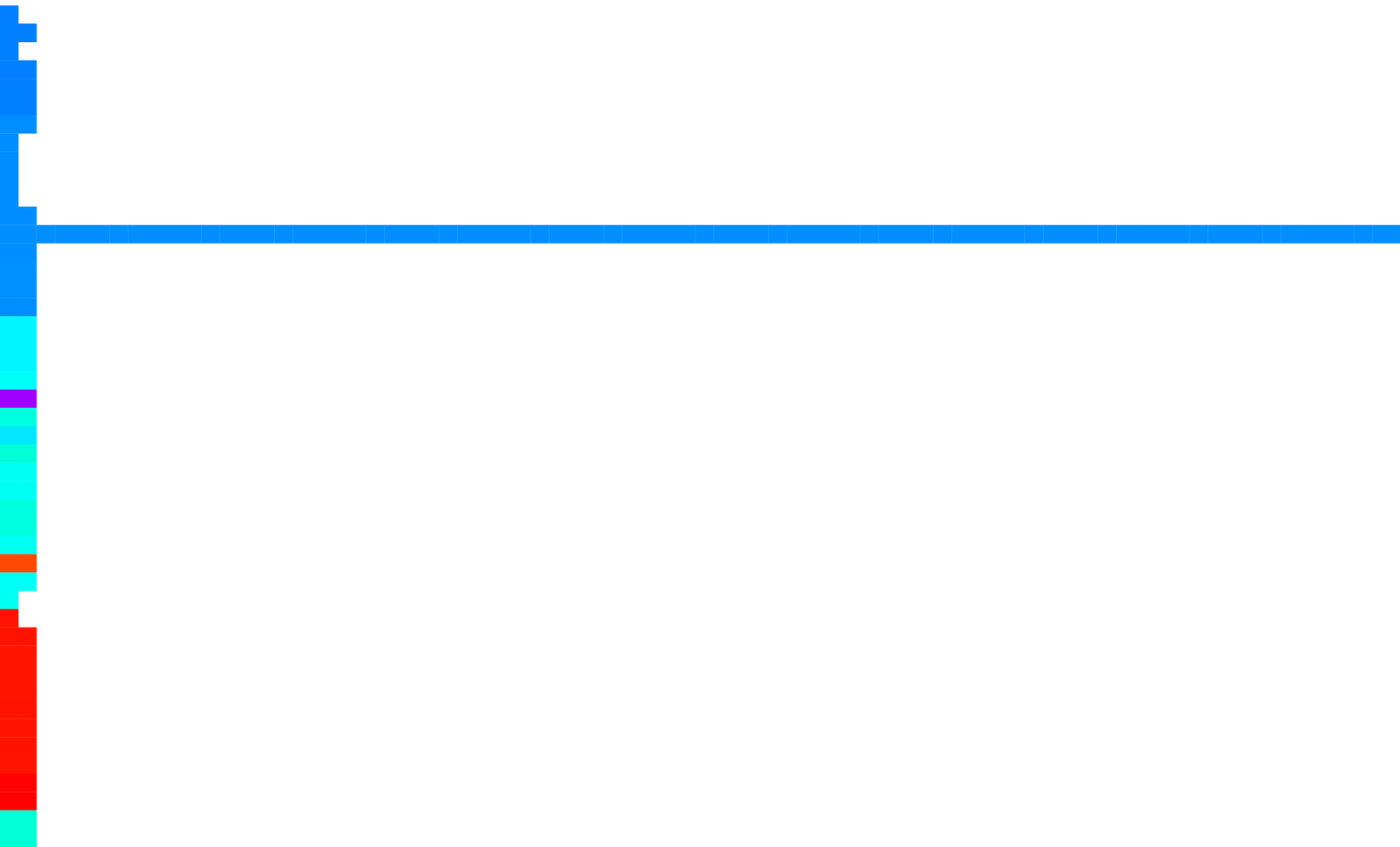}
}
\end{tabular}
\end{center}
\caption{Visualization of reference (a) and reconstructed (bcd) alignments 
for test case \textsc{bb40037.}
In all alignments sequences are ordered accordingly.
Each sequence is colored based on the evolutionary distance to its
neighbors in a phylogenetic tree, such that families
of related sequences have similar colors. Trees for (a) and (b)
are computed with the PhyML 3.0 program \cite{Guindon2010},
using the maximum parsimony method. Trees for (c) and (d) are
the guide-trees used by those aligners.\label{fig:test-bb40037-visualisation}}
\end{figure}

\begin{figure}[!t]
\begin{tabular}{lc}
{\includegraphics[width=0.25\columnwidth,height=0.2\columnwidth]{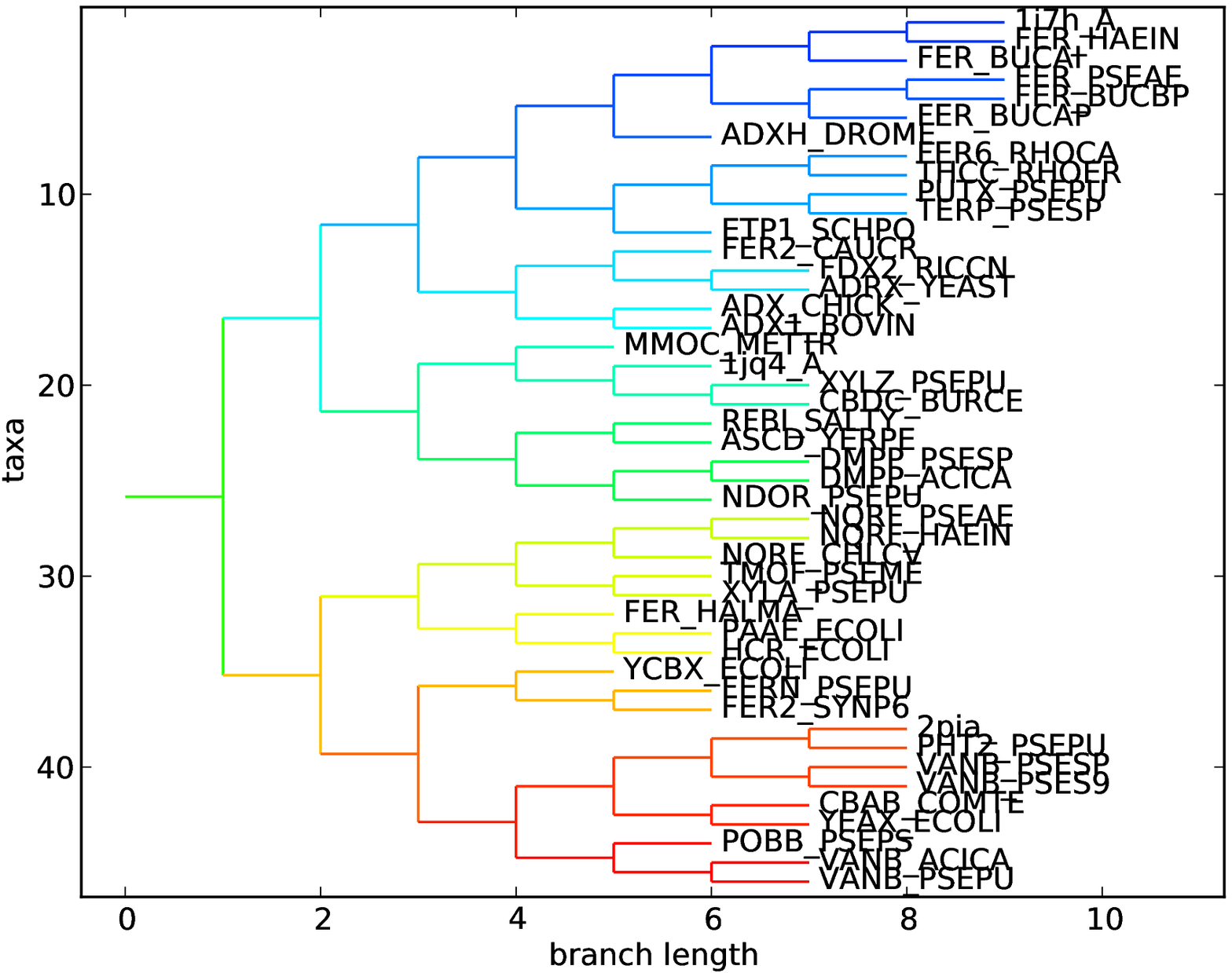}}
&
{\includegraphics[width=0.6\columnwidth,height=0.2\columnwidth]{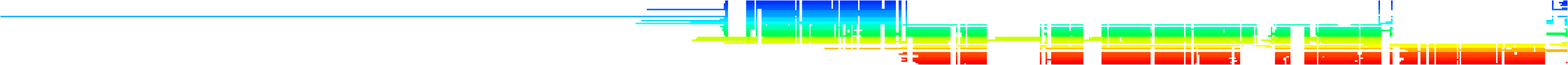}
}\\
(a) & Probalign\hspace{30ex}(b)\\\\
{\includegraphics[width=0.25\columnwidth,height=0.2\columnwidth]{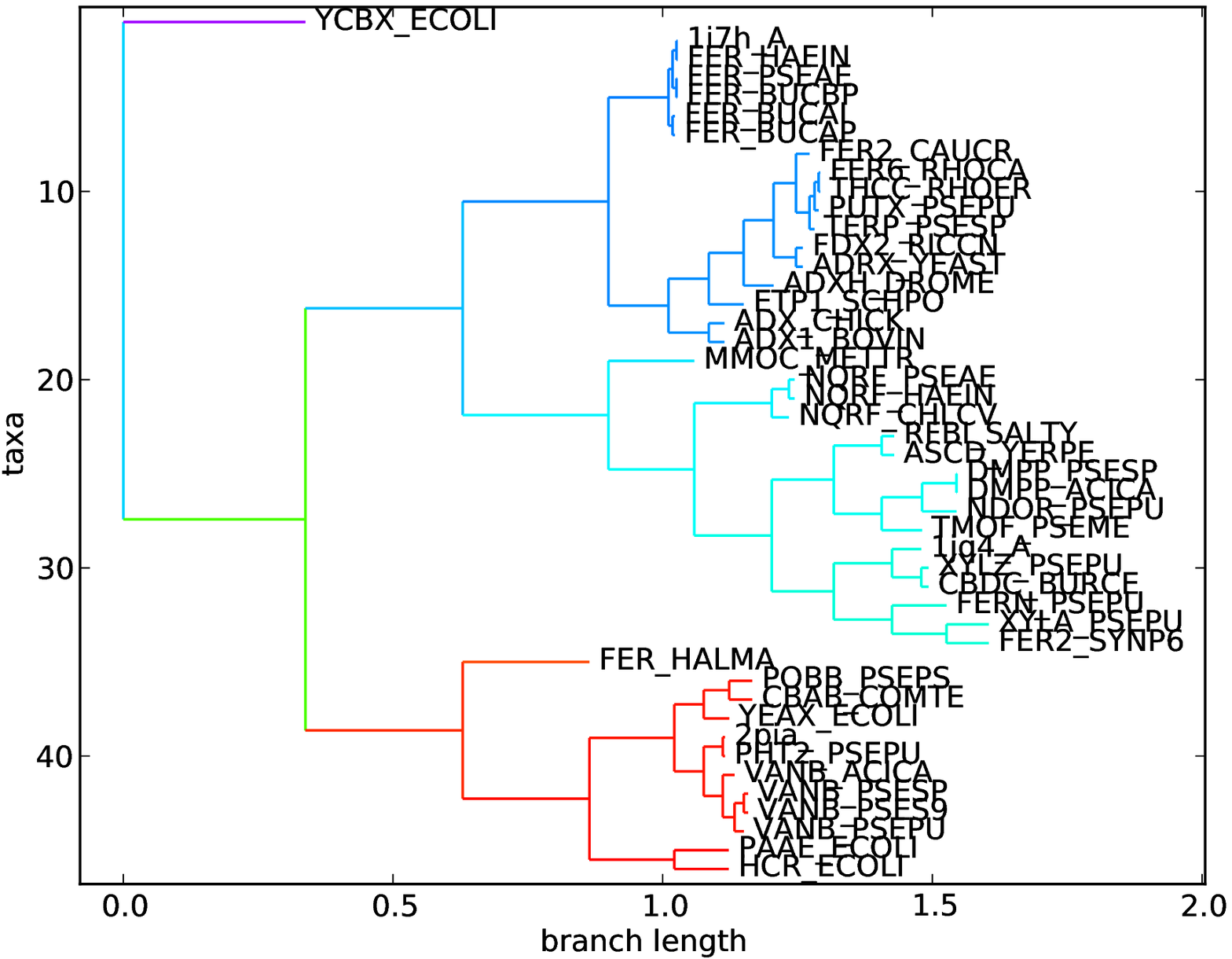}}
&
{\includegraphics[width=0.7\columnwidth,height=0.2\columnwidth]{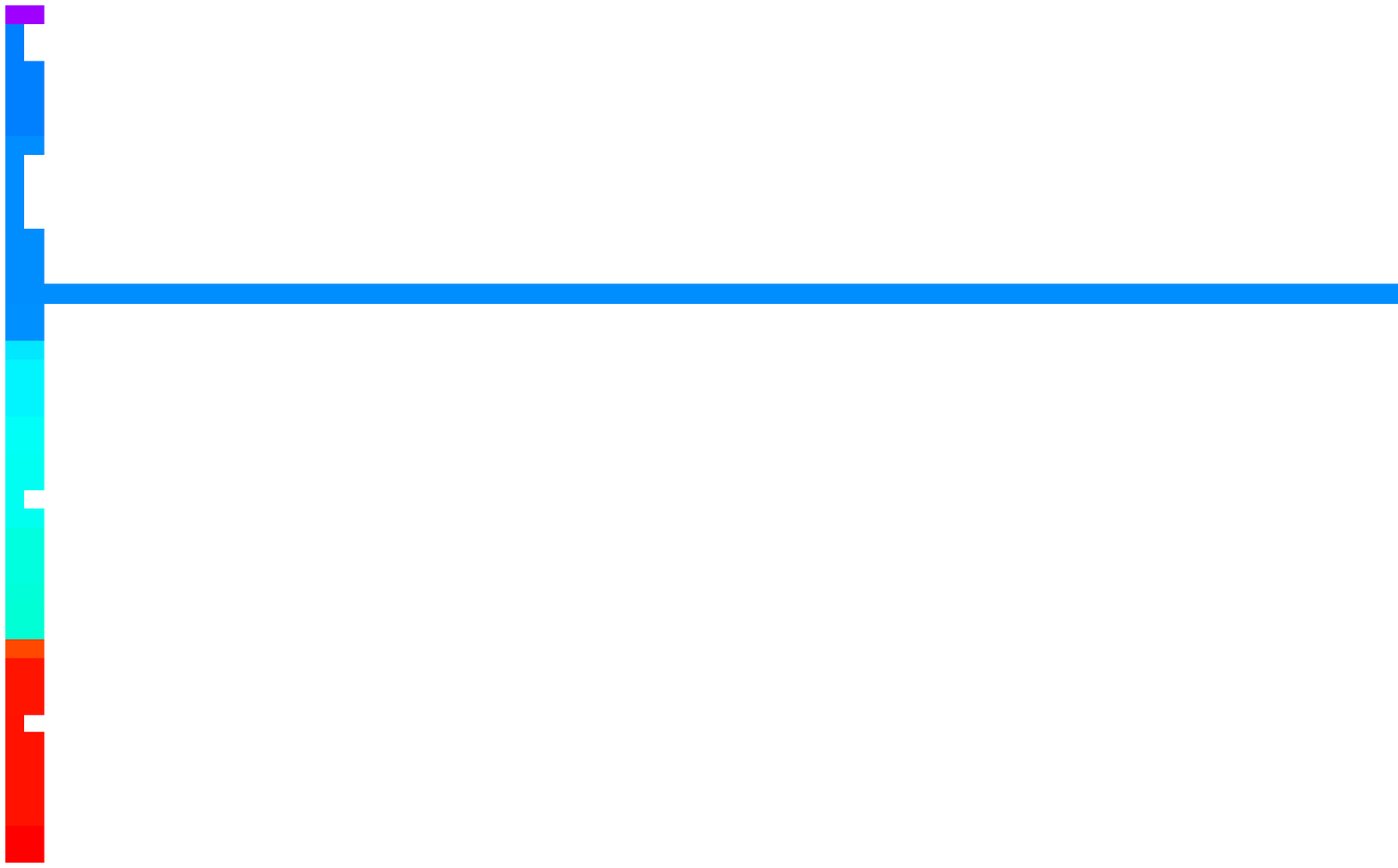}}\\
(c) & MSAProbs\hspace{30ex}(d)
\end{tabular}

\caption{Guide trees (ac) and alignment visualizations (bd) for test case \textsc{bb40037} and programs Probalign (ab) and MSAProbs (cd).
Tree branches and aligned sequences are colored 
based on the evolutionary distances to their neighbors, 
as computed from the guide-trees used during alignment.
Sequences in alignments are ordered following their order in trees,
so related sequences have similar color and are positioned together.
\label{fig:test-bb40037-reordered}}
\end{figure}

We illustrate this observation with an example of test case \textsc{bb40037}. 
As is shown in column 9 of Table \ref{tab:results},
MSARC outperforms other methods by a large margin. 
The TC scores of zero means that each alignment method has shifted
at least one sequence from its correct position relative to the other
sequences. 
Figure \ref{fig:test-bb40037-visualisation} presents the
structure of the reference alignment, as well as
alignments generated by MSARC, Probalign and MSAprobs. 
The large family of red, orange and yellow colored sequences near the bottom
has been misaligned by the progressive methods. 
The reason for this is more visible 
in Figure \ref{fig:test-bb40037-reordered},
where sequences in alignments are reordered
according to related guide-trees.

Probalign aligns separately the first half of the sequences (blue and green) 
and the second half of the sequences (from yellow to red).
Next, the prefixes of the
second group are aligned with the suffixes of the first group,
propagating an error within a yellow sub-alignment%
.

MSAprobs aligns separately the dark blue, light blue and red sequences. 
Next the blue sub-alignments are aligned together. 
\nowe{Resulting alignment has erroneously inserted gaps near the right ends
of dark blue sequences.}
\nowe{This error is propagated in next step, 
where the suffix of the blue alignment 
is aligned with the prefix of the red alignment.}
Finally the single violet sequence %
is added to the alignment, splitting it in two.

\nowe{For both programs, alignment errors introduced in the earlier steps 
are propagated to the final alignment. 
On the other hand, 
the non-progressive strategy used in MSARC 
yields a reasonable approximation of the reference alignment
(see Figure \ref{fig:test-bb40037-visualisation}(ab)).}

\section{Discussion}



The progressive principle dominates multiple alignment algorithms for nearly 20 years.
Throughout this time, many groups have dedicated their effort 
to refine its accuracy to the current state.
Other approaches were omitted due to high computational complexity and/or
unsatisfactory quality.
To our best knowledge, MSARC is the only non-progressive aligner
of quality comparable to best progressive programs.
Moreover, due to a guide-tree bias of alignments computed with progressive methods,
MSARC is a quality leader for sequence sets 
with evolutionary distances hardly representable by a phylogenetic tree.

Despite of the algorithmic novelty, 
the non-progressive approach to multiple alignment
makes MSARC an interesting tool for phylogeny reconstruction pipelines.
The objective of these procedures is to infer the structure of 
a phylogenetic tree from a given sequence set.
Multiple alignment is usually the first pipeline step.
When alignment is guided by a tree, 
the reconstructed phylogeny is biased towards this tree.
In order to minimize this effect, 
some phylogenetic pipelines alternately 
optimize a tree and an alignment
\cite{Redelings2005,Lunter2005,Liu2009}.
Unbiased alignment process of MSARC may simplify this procedure
and improve the reconstruction accuracy, 
especially in most problematic cases.

The main disadvantage of MSARC is its computational complexity,
especially in the case of the multilevel scheme variant
(MSARC-FM is $\sim 3\times$ slower than MSAProbs 
and $\sim 5\times$ slower than Probalign,
MSARC-ML is $1.5\times$ slower than MSARC-FM).
However, the running time can be greatly
improved by using multiple cores to parallel computations,
because every step of its algorithm can be parallelized.
Since multiple cores are becoming more and more common, this should
allow for the computation time comparable with other alignment algorithms.

MSARC has also the potential for quality improvements. 
Alternative methods of computing residue alignment affinities 
could be used to improve the accuracy of both MSARC and Probalign based methods. 
Other approaches to alignment graph partitioning may also lead 
to improvements in the accuracy of MSARC, 
for example a better method of pairing residues for multilevel coarsening 
than currently used naive consecutive neighbors merging.

\subsubsection*{Acknowledgements}
This work was supported by the Polish Ministry of Science and Higher Education 
[N N519 652740].

\bibliographystyle{splncs03}
\bibliography{mulign}

\begin{thebibliography}{10}
\providecommand{\url}[1]{\texttt{#1}}
\providecommand{\urlprefix}{URL }

\bibitem{Do2005}
Do, C.B., Mahabhashyam, M.S.P., Brudno, M., Batzoglou, S.: Probcons:
  Probabilistic consistency-based multiple sequence alignment. Genome Res
  15(2),  330--340 (Feb 2005), \url{http://dx.doi.org/10.1101/gr.2821705}

\bibitem{Edgar2004a}
Edgar, R.C.: Muscle: multiple sequence alignment with high accuracy and high
  throughput. Nucleic Acids Res  32(5),  1792--1797 (2004),
  \url{http://dx.doi.org/10.1093/nar/gkh340}

\bibitem{Fiduccia1982}
Fiduccia, C.M., Mattheyses, R.M.: A linear-time heuristic for improving network
  partitions. In: Proceedings of the 19th Design Automation Conference. pp.
  175--181. DAC '82, IEEE Press, Piscataway, NJ, USA (1982),
  \url{http://dl.acm.org/citation.cfm?id=800263.809204}

\bibitem{Gonnet1992}
Gonnet, G.H., Cohen, M.A., Benner, S.A.: Exhaustive matching of the entire
  protein sequence database. Science  256(5062),  1443--1445 (Jun 1992)

\bibitem{Gotoh1982}
Gotoh, O.: An improved algorithm for matching biological sequences. J Mol Biol
  162(3),  705--708 (Dec 1982)

\bibitem{Guindon2010}
Guindon, S., Dufayard, J.F., Lefort, V., Anisimova, M., Hordijk, W., Gascuel,
  O.: New algorithms and methods to estimate maximum-likelihood phylogenies:
  assessing the performance of phyml 3.0. Syst Biol  59(3),  307--321 (May
  2010), \url{http://dx.doi.org/10.1093/sysbio/syq010}

\bibitem{Hendrickson1995}
Hendrickson, B., Leland, R.: A multilevel algorithm for partitioning graphs.
  In: Proceedings of the 1995 ACM/IEEE conference on Supercomputing (CDROM).
  Supercomputing '95, ACM, New York, NY, USA (1995),
  \url{http://doi.acm.org/10.1145/224170.224228}

\bibitem{Katoh2005a}
Katoh, K., ichi Kuma, K., Toh, H., Miyata, T.: Mafft version 5: improvement in
  accuracy of multiple sequence alignment. Nucleic Acids Res  33(2),  511--518
  (2005), \url{http://dx.doi.org/10.1093/nar/gki198}

\bibitem{Kececioglu1993}
Kececioglu, J.: The maximum weight trace problem in multiple sequence
  alignment. In: Proceedings of the 4th Symposium on Combinatorial Pattern
  Matching (CPM). pp. 106--119. No. 684 in Lecture Notes in Computer Science,
  Springer-Verlag (1993)

\bibitem{Liu2009}
Liu, K., Raghavan, S., Nelesen, S., Linder, C.R., Warnow, T.: Rapid and
  accurate large-scale coestimation of sequence alignments and phylogenetic
  trees. Science  324(5934),  1561--1564 (Jun 2009),
  \url{http://dx.doi.org/10.1126/science.1171243}

\bibitem{Loeytynoja2008a}
L{\"o}ytynoja, A., Goldman, N.: Phylogeny-aware gap placement prevents errors
  in sequence alignment and evolutionary analysis. Science  320(5883),
  1632--1635 (Jun 2008), \url{http://dx.doi.org/10.1126/science.1158395}

\bibitem{Lunter2005}
Lunter, G., Miklós, I., Drummond, A., Jensen, J.L., Hein, J.: Bayesian
  coestimation of phylogeny and sequence alignment. BMC Bioinformatics  6, ~83
  (2005), \url{http://dx.doi.org/10.1186/1471-2105-6-83}

\bibitem{Miyazawa1995}
Miyazawa, S.: A reliable sequence alignment method based on probabilities of
  residue correspondences. Protein Eng  8(10),  999--1009 (Oct 1995)

\bibitem{Mueckstein2002}
M{\"u}ckstein, U., Hofacker, I.L., Stadler, P.F.: Stochastic pairwise
  alignments. Bioinformatics  18 Suppl 2,  S153--S160 (2002)

\bibitem{Notredame2000}
Notredame, C., Higgins, D.G., Heringa, J.: T-coffee: A novel method for fast
  and accurate multiple sequence alignment. J Mol Biol  302(1),  205--217 (Sep
  2000), \url{http://dx.doi.org/10.1006/jmbi.2000.4042}

\bibitem{Redelings2005}
Redelings, B.D., Suchard, M.A.: Joint bayesian estimation of alignment and
  phylogeny. Syst Biol  54(3),  401--418 (Jun 2005),
  \url{http://dx.doi.org/10.1080/10635150590947041}

\bibitem{Roshan2006}
Roshan, U., Livesay, D.R.: Probalign: multiple sequence alignment using
  partition function posterior probabilities. Bioinformatics  22(22),
  2715--2721 (Nov 2006), \url{http://dx.doi.org/10.1093/bioinformatics/btl472}

\bibitem{Sievers2011}
Sievers, F., Wilm, A., Dineen, D., Gibson, T.J., Karplus, K., Li, W., Lopez,
  R., McWilliam, H., Remmert, M., Söding, J., Thompson, J.D., Higgins, D.G.:
  Fast, scalable generation of high-quality protein multiple sequence
  alignments using clustal omega. Mol Syst Biol  7,  539 (2011),
  \url{http://dx.doi.org/10.1038/msb.2011.75}

\bibitem{Subramanian2008}
Subramanian, A.R., Kaufmann, M., Morgenstern, B.: Dialign-tx: greedy and
  progressive approaches for segment-based multiple sequence alignment.
  Algorithms Mol Biol  3, ~6 (2008),
  \url{http://dx.doi.org/10.1186/1748-7188-3-6}

\bibitem{Subramanian2005}
Subramanian, A.R., Weyer-Menkhoff, J., Kaufmann, M., Morgenstern, B.:
  Dialign-t: an improved algorithm for segment-based multiple sequence
  alignment. BMC Bioinformatics  6, ~66 (2005),
  \url{http://dx.doi.org/10.1186/1471-2105-6-66}

\bibitem{Thompson1994}
Thompson, J.D., Higgins, D.G., Gibson, T.J.: Clustal w: improving the
  sensitivity of progressive multiple sequence alignment through sequence
  weighting, position-specific gap penalties and weight matrix choice. Nucleic
  Acids Res  22(22),  4673--4680 (Nov 1994)

\bibitem{Thompson2005}
Thompson, J.D., Koehl, P., Ripp, R., Poch, O.: Balibase 3.0: latest
  developments of the multiple sequence alignment benchmark. Proteins  61(1),
  127--136 (Oct 2005), \url{http://dx.doi.org/10.1002/prot.20527}

\bibitem{Wong2008}
Wong, K.M., Suchard, M.A., Huelsenbeck, J.P.: Alignment uncertainty and genomic
  analysis. Science  319(5862),  473--476 (Jan 2008),
  \url{http://dx.doi.org/10.1126/science.1151532}

\bibitem{Yu2001}
Yu, Y.K., Hwa, T.: Statistical significance of probabilistic sequence alignment
  and related local hidden markov models. J Comput Biol  8(3),  249--282
  (2001), \url{http://dx.doi.org/10.1089/10665270152530845}

\end{thebibliography}

\end{document}